\documentstyle[12pt, aaspp4,flushrt]{article}

\newcommand{\Ref}{\hangindent=20pt \hangafter=1 \noindent}
\newcommand{\StartRef}{\hyphenpenalty=10000 \raggedright}

\newcommand{\beq}{\begin{equation}}
\newcommand{\eeq}{\end{equation}}
\newcommand{\NarrowMargins}{
  \setlength{\oddsidemargin}{+0.3in}
  \setlength{\evensidemargin}{-0.0in}
  \setlength{\textwidth}{6.2in}
  \setlength{\topmargin}{-0.75in}
  \setlength{\textheight}{9.25in}   }
\catcode`\@=11 

\def\lsim{\mathrel{\mathpalette\@versim<}}
\def\gsim{\mathrel{\mathpalette\@versim>}}
\def\vp{v_{\parallel}}

\def\om{\omega}

\def\op{\Omega_p}
\def\rp{\rho_p}
\def\kp{k_{\perp}}
\def\kpa{k_{\parallel}}

\def\omnl{\omega_{nl}}

\def\@versim#1#2{\vcenter{\offinterlineskip
        \ialign{$\m@th#1\hfil##\hfil$\crcr#2\crcr\sim\crcr } }}
\catcode`\@=12 
\NarrowMargins
\input{psfig}

\begin{document}
\title{Turbulence and Particle Heating in Advection-Dominated
Accretion Flows}
\author{ Eliot Quataert$^{\dagger}$ and Andrei Gruzinov$^{\dagger \dagger}$ }
\affil{$^{\dagger}$ Harvard-Smithsonian Center for Astrophysics, 60 Garden 
Street, Cambridge, MA 02138; equataert@cfa.harvard.edu}
\affil{$^{\dagger \dagger}$ Institute for Advanced Study, School of Natural 
Sciences,
Princeton, NJ 08540; andrei@ias.edu}

\begin{abstract}

We extend and reconcile recent work on turbulence and particle heating
in advection-dominated accretion flows.  For approximately
equipartition magnetic fields, the turbulence primarily heats the
electrons.  For weaker magnetic fields, the protons are primarily
heated.  The division between electron and proton heating occurs
between $\beta \sim 5$ and $\beta \sim 100$ ($\beta \equiv$ ratio of
gas to magnetic pressure), depending on unknown details of how Alfv\'en
waves are converted into whistlers on scales of the proton Larmor
radius.  We also discuss the possibility that magnetic reconnection
could be a significant source of electron heating.

\noindent {\em Subject headings:} accretion -- hydromagnetics -- plasmas
-- turbulence

\end{abstract}

\section{Introduction}
Phenomenological models of accretion flows should predict a radiation
spectrum for a central object of a given mass using a few relevant
dimensionless parameters (e.g., the mass accretion rate in Eddington
units, the Shakura-Sunyaev viscosity parameter, $\alpha$, and the
ratio of the gas to the magnetic pressure, $\beta$). In
advection-dominated accretion flows (ADAFs), which are hot
collisionless plasmas that have been argued to form around some
accreting black holes, an additional dimensionless parameter, the
fraction of the turbulent energy which heats the electrons ($\equiv
\delta$), becomes important.  In this paper, we discuss the electron
heating mechanisms in ADAFs; in particular, we show that $\delta$
should grow as $\beta$ decreases.

In ADAFs (Ichimaru 1977; Rees et al. 1982; Narayan \& Yi 1994, 1995;
Abramowicz et al. 1995), the accreting gas is unable to cool
efficiently and most of the energy generated by viscous stresses is
stored as thermal energy of the gas and is advected onto the central
object.  As a result, the gas heats up to nearly virial temperatures
and adopts a two-temperature configuration, with the protons
significantly hotter than the radiating electrons (Rees et al. 1982;
Narayan \& Yi 1995).

In the context of ADAFs, the issue of which particles (electrons or
protons) receive the viscous energy acquires particular importance.
In ``standard'' ADAF models (see Narayan et al. 1998 for a review)
inefficient cooling of the gas occurs because the viscously generated
energy is assumed to primarily heat the protons.  Since the accretion
flow is effectively collisionless (e.g., Rees et al. 1982) only a
small fraction of this energy is transferred to the electrons via
Coulomb collisions; the total energy radiated by the gas (almost all
by the electrons) is therefore much less than the total energy
generated by viscosity.  If viscosity were to predominantly heat the
electrons, an accretion flow could be advection-dominated only at very
low densities (and thus very low accretion rates) when the electrons
themselves are unable to cool efficiently.  As a result, the optically
thin ADAF formalism would probably not be relevant for observed
systems.

Recently there has been some theoretical progress in addressing the
issue of particle heating in ADAFs.  The basic physical picture is
that ADAFs, like thin disks, are magnetized turbulent plasmas.  On
small scales (much less than the outer scale of the turbulence, which
is $\sim$ the local radius in the accretion flow), the turbulence will
be approximately incompressible.  Since the Alfv\'en wave is the
incompressible MHD mode, turbulence on relatively small scales can be
described as a spectrum of nonlinearly interacting Alfv\'en waves.
These nonlinear interactions transfer energy from large scale
perturbations (waves) to smaller scales until it is dissipated.  The
relative heating of protons and electrons is determined by which
particle species is responsible for dissipating the turbulence.

There are (at least) two physical processes which are neglected by
simply equating particle heating in ADAFs with particle heating by
Alfv\'enic turbulence.  The first (which, in our view, is potentially
important) is magnetic reconnection, which may occur at energetically
important levels in the accretion flow.  This is discussed in \S5 and
in Appendix B.  

The second (which, in our view, is less important) is that, on large
scales, the turbulence may have a compressible component.  This
possibility has been considered (implicitly) by Blackman (1998), who
argues that Fermi acceleration by relatively large scale (much larger
than $\rp \equiv$ the Larmor radius of thermal protons) magnetic
perturbations associated with MHD turbulence preferentially heats the
protons to the degree required by ADAF models.  This conclusion is,
however, equivalent to considering the collisionless dissipation of
the fast mode component of MHD turbulence (which is compressible), and
thus does not apply to the entire turbulent cascade (Achterberg 1981).
The incompressible component of MHD turbulence, i.e., the Alfv\'en
waves, is undamped on spatial scales much larger than $\rp$ (see \S2).

Two qualitative considerations suggest to us that compressible
turbulence may be energetically less important than Alfv\'enic
turbulence: (1) the Balbus-Hawley instability, which is likely
relevant for the generation of magnetic fields and turbulence in the
accretion flow, is non-compressive to linear order (Balbus \& Hawley
1991).  (2) if the turbulent velocity is subsonic, the generation of
compressive turbulence may be severely suppressed.  Throughout this
paper, we therefore focus exclusively on Alfv\'enic turbulence.  Our
results can, of course, be trivially recast in terms of the fraction
(likely $\sim 1$) of the viscously generated energy residing in
Alfv\'enic turbulence.

Gruzinov (1998; hereafter G98) and Quataert (1998; hereafter Q98)
analyzed particle heating by the Alfv\'enic component of MHD turbulence.
In this paper we provide a synthesis and extension of their work.  We
first review Alfv\'enic turbulence and wave dissipation on length scales
comparable to or larger than $\rp$ (\S2).  The primary difference
between Gruzinov and Quataert's analysis is that Quataert's was
restricted to these ``large'' length scales, while Gruzinov included a
preliminary analysis of what happens on smaller scales.  In \S3 (and
Appendix A) we provide a more detailed discussion of turbulence and
wave dissipation on length scales smaller than $\rp$. We then present
a model for how the turbulent energy cascades past $\rp$ (\S4).  The
principle results of this paper are given in \S5.  This contains our
best estimate, along with what we take to be plausible uncertainties,
of the electron heating rate in ADAFs.

\section{Kolmogorov-Goldreich-Sridhar Turbulence}
To investigate particle heating by turbulence, we must understand both
the dynamics of the turbulent cascade, as well as the dissipation
mechanisms which operate on it.  Recent work by Goldreich \& Sridhar
(1995; hereafter GS) provides the necessary characteristics of
Alfv\'enic turbulence.  In the MHD limit, linear Alfv\'en waves satisfy
the dispersion relation $\omega = v_A |\kpa|$, where $\omega$ is the
mode frequency, $v_A$ is the Alfv\'en speed and $\kpa$ is the
component of the wavevector along the mean magnetic field.

GS argue (and recent numerical calculations confirm; Maron 1998) that
Alfv\'enic turbulence naturally evolves into a critically balanced
state in which the timescale for nonlinear effects to transfer energy
from a wavevector $\sim \bf k$ to a wavevector $\sim 2 \bf k$
($\equiv$ the cascade time, $T_c$) is comparable to the linear wave
period at that scale, $T \equiv 2 \pi / \omega$; this determines how
rapid the dissipation must be to halt the cascade.  It also implies
that the cascade is highly anisotropic, with the energy cascading
primarily perpendicular to the local magnetic field; the parallel and
perpendicular sizes of a wave at any scale are correlated, with $\kpa
\sim k^{2/3}_{\perp} R^{-1/3}$, where $R$, the outer scale of the
turbulence, is $\sim$ the local radius in the accretion flow. Since
the dissipation of an Alfv\'en wave depends on its direction with
respect to the background magnetic field (see below), the path of the
cascade in wavevector space is crucial.

In collisionless plasmas, such as those in ADAFs, the wave dissipation
mechanisms of principle importance are wave-particle resonances
(molecular viscosity, thermal conductivity, electrical resistivity,
etc. are entirely unimportant). Resonance occurs when the frequency of
the wave, in the frame moving with the particle along the field line,
is an integer multiple of the particle's cyclotron frequency
($\Omega$), \beq \omega - \kpa \vp = n \Omega, \label{res} \eeq where
$\vp$ is the particle's velocity along the magnetic field.  For $n =
0$, resonance occurs when the wave's phase speed along the field line,
$v = \omega/\kpa$, equals $\vp$.  A necessary (but not
sufficient) condition for strong damping is that $v$ be comparable
to the thermal speed of the particles, so that there are a large
number of resonant particles.

The $n = 0$ resonance corresponds to two physically distinct
wave-particle interactions. In Landau damping (LD), particle
acceleration is due to the wave's longitudinal electric field
perturbation (i.e., the usual electrostatic force, $E_z$).  The ratio
of the electron to the proton heating rates ($\equiv P$) for a wave
(with $\kp \rp \lsim 1$) damped solely by LD is $P_{LD} \approx (m_e
T^3_p/m_p T_e^3)^{1/2}$, which is $\gg 1$ for the $T_p \gg T_e$
plasmas of interest to us.  In transit-time damping (TTD), the
magnetic analogue of LD, the interaction is between the particle's
effective magnetic moment ($\mu = m v^2_{\perp}/2 B$) and the wave's
longitudinal magnetic field perturbation, $B_z$ (Stix 1992).  For a
wave (with $\kp \rp \lsim 1$) damped solely by TTD, the protons are
preferentially heated: $P_{TTD} \approx (m_e T_e/m_p T_p)^{1/2} \ll
1$.\footnote{In this and the previous expression we have taken $v
\lsim$ the electron and proton thermal speeds; see Q98 for a more
general expression.}  This is because in plasmas with $T_p \gg T_e$,
the protons have the larger magnetic moment and so couple better to a
wave's magnetic field perturbation.

The Alfv\'en wave has $|v| = v_A$; in plasmas appropriate to ADAFs
($T_p \gg T_e$ and $\beta \gsim 1$, where $\beta$ is the ratio of the
gas pressure to the magnetic pressure), $v_A$ is comparable to the
electron and proton thermal speeds and so there are a large number of
particles available to resonate with the wave.  In the MHD limit,
however, the Alfv\'en wave has both $E_z = 0$ and $B_z = 0$ and so is
undamped by linear collisionless effects.\footnote{By contrast, the
fast mode has $B_z \ne 0$ in the MHD limit and is thus strongly damped
by TTD.  This is why protons are preferentially heated by compressive
MHD turbulence in ADAFs.}  For larger wavevectors, the MHD
approximations are less applicable and kinetic theory corrections to
$E_z$ and $B_z$ become important, leading to finite dissipation.  For
the perpendicular cascade of Alfv\'en waves due to GS, however, when
$\kp \rp \sim 1$, $\om \sim \op(\rp/ R)^{1/3} \beta^{-1/2} \ll \op$
and so $n \ne 0$ resonances can be satisfied only by particles with
$|\vp| \sim |n \Omega| v_A/\omega \gg v_A$, of which there are a
negligible number.  The cyclotron resonance is thus unimportant in
dissipating the turbulent energy in the GS cascade.

In order to accurately assess the properties of Alfv\'en waves with $\kp
\sim \rp^{-1} \gg \kpa$, we have solved the full kinetic theory
dispersion relation for linear perturbations to a warm plasma. These
calculations are utilized throughout this paper, but we refer the
reader to G98 or Q98 for details.  Here, and in Figure 1, we summarize
the primary results.  Alfv\'en waves with $\kp \sim \rp^{-1} \gg \kpa$
are damped primarily by the protons (via TTD).  The dissipation rate,
$\gamma$, is essentially independent of $T_p/T_e$, but is an
increasing function of $\beta$ ($\gamma/\omega \approx 0.16 \beta^{1/2} (\kp
\rp)^2$ for $\beta \gg 1$ and $\kp \rho_p \lsim 1$).  For plasmas
with $\beta \gg 1$, Alfv\'enic turbulence is therefore dissipated on
length scales $\gsim \rp$, with most of the turbulent energy heating
the protons (this is quantified in \S5).

For $\beta = 1$, however, the maximal dissipation rate for an Alfv\'en
wave in the GS cascade (obtained at $\kp \rp = 1$) is $\gamma T
\approx 0.1$ (see Figure 1).  Since the timescale for energy to
cascade through the inertial range is $T_c \approx T$, this suggests
that, for plasmas with equipartition magnetic fields ($\beta \approx
1$), very little of the turbulent energy is dissipated on scales
comparable to or greater than $\rp$.  Consequently, we must
investigate the flux of turbulent energy past $\kp \rho_p \approx 1$,
as well as turbulence and wave dissipation on scales smaller than
$\rp$.
 
\section{Whistler Turbulence}
Alfv\'en waves only exist for $\kp \rp \lsim 1$.  For $\kp \rp \gsim 1$,
the same mode is called the whistler.  In Appendix A we show that
all of the turbulent energy which is not damped on scales of $\kp \rp
\lsim 1$ is transformed into whistlers at $\kp \rp \gsim 1$ (in
particular, we argue that there are no other channels through which
the energy can travel).

The following argument regarding particle heating by whistler
turbulence can be provided.  For $\kp \rp \gsim 1$, but $\kp \gg
\kpa$, whistlers have $\omega \ll \Omega_p$.  Thus, in a mode period,
a particle undergoes many Larmor orbits.  Since $\kp \rp \gsim 1$, the
protons (but not the electrons) sample a rapidly varying
electro-magnetic field in the course of a Larmor orbit.  As a result,
the protons are ``frozen out'' and become dynamically unimportant
(they provide a uniform background of positive charge).  In
particular, they cannot contribute to damping the whistler energy.
Whistler energy therefore cascades to smaller length scales until it
is damped by the electrons.

The more detailed analysis of whistler turbulence and particle heating
given in Appendix A confirms this picture.  The result hinges
crucially on our estimate that whistler turbulence maintains $\kpa \ll
\kp$ (recall that the GS cascade gives $\kpa \sim \kp (\rp/R)^{1/3}
\sim 10^{-3} \kp$ at $\kp \rp \sim 1$).  In this case Larmor circle
averaging is efficient and the whistler energy is dissipated by the
electrons (via Landau damping) on scales of $\kp \rp \sim 30$ (see
Fig. 1).  If whistler turbulence were to reach the proton cyclotron
frequency before $\kp \rp \sim 10$ (which we argue is unlikely), the
whistler energy would heat the protons via the cyclotron resonance
(see Appendix A.3).

\section{Energy Flux through the Damping Barrier}
We therefore assume that most of the energy dissipated on scales
$\gsim \rp$ heats the protons, while the energy that gets past $\kp
\rp \sim 1$ heats the electrons.  Here we give a crude model for
calculating the damping of the turbulent energy flux; this provides an
estimate of the fraction of the energy flux that gets through the
proton damping barrier at $\kp \rp \sim 1$, which is needed to
calculate the electron heating rate ($\delta$).

We define the full energy spectrum of the turbulence (normalized to
the plasma density) as $E(\kp, \kpa)$ and the perpendicular energy
spectrum as $E(\kp$):
\begin{equation}
<v^2>=\int d \kp d \kpa E(k_{\perp}, \kpa) = \int d \kp E(\kp),
\end{equation}
where $v$ is the plasma velocity.  As we are interested in the flux of
energy past $\kp \rp = 1$, we only need $E(\kp)$ for $\kp \rp \lsim
1$, which is given by the Kolmogorov-Goldreich-Sridhar spectrum
\begin{equation}
E(k_{\perp})= C_1 \epsilon ^{2/3} k_{\perp}^{-5/3}, \label{ekp}
\end{equation}
where $\epsilon ={\rm cm^2/s^3}$ is the energy flux in wavenumber
space and $C_1$ is a dimensionless constant of order unity (a
Kolmogorov constant).  There is essentially no dissipation at
$k_{\perp}\rp \ll 1$, and so $\epsilon$ is constant ($\equiv
\epsilon_0$) in this region.  At larger $k_{\perp}$, the energy flux
$\epsilon$ is a monotonically decreasing function of $k_{\perp}$. The
decrease of the energy flux is described by
\begin{equation}
{d\epsilon \over dk_{\perp}}= -2 <\gamma(\kp) > E(\kp),
\end{equation}
where $<\gamma(\kp) >$ is the parallel wave number averaged damping rate,
\begin{equation}
<\gamma(\kp) >={\int dk_{\parallel}\gamma 
(k_{\perp},k_{\parallel})E(k_{\perp},k_{\parallel}) \over E(k_{\perp}) }.
\end{equation}

For the GS cascade, the turbulent energy lies inside the cone $\kpa
\lsim \omnl/v_A$, where $\omnl = C_2 \epsilon^{1/3} \kp^{2/3}$ is the
nonlinear frequency of the turbulence (at scale $\kp$) and $C_2$ is
another dimensionless constant of order unity.  We model this using a
simple expression for $E(\kp,\kpa)$, namely, \beq E(\kp,\kpa) = E(\kp)
\delta(\kpa - \omnl/v_A). \label{ekpa} \eeq This corresponds to all of
the turbulent energy flowing along the line in $k$ space for which the
linear mode frequency ($\omega = v_A |\kpa|$) is equal to the
nonlinear frequency of the turbulence (which captures the key physical
result of the GS cascade).  Using equations (\ref{ekp})-(\ref{ekpa}),
it is straightforward to show that \beq \epsilon = \epsilon_0 \exp
\left[ -2 C \int d \log \kp D(\kp) \right],
\label{flux} \eeq where $D(\kp) \equiv
\gamma(\kp,\kpa)/\omega(\kp,\kpa)$ (with $\kpa = \omnl/v_A$) is the
dimensionless damping rate of the mode (the inverse of the quality
factor; see Fig. 1) and $C = C_1 C_2$.

Recent numerical simulations confirm the GS picture of Alfv\'enic
turbulence (Maron 1998).  They also give values for the dimensionless
constants in the above expressions, namely $C_1 = 2.5 \pm 0.6$ and
$C_2 = 2.2 \pm 0.4$, so that $C \approx 6$.

We are interested in the energy flux past $\kp \rp \sim 1$.  Since $D
\propto \kp^2$, the integral in equation (\ref{flux}) is dominated by
the contribution from $\kp \rp \sim 1$, which lies outside the region
where the MHD simulations are valid.  Consequently, $C \approx 6$ must
be taken as a crude approximation, subject to significant
uncertainties.  The basic uncertainty is that we do not understand the
details of how Alfv\'enic turbulence is converted into whistler
turbulence on scales of $\rp$.  For example, Alfv\'en waves with $\kp
\rp \lsim 1$ can excite whistlers with $\kp \rp \gsim 1$ by three wave
interactions (see Appendix A). However, the effective $\kp \rp$ at
which this occurs is uncertain.  Consequently, so is the upper limit
for the integral in equation (\ref{flux}).  This uncertainty can be
absorbed into an uncertainty in C.  In addition, given that Alfv\'en
waves excite whistlers, the precise timescale on which this occurs is
uncertain (i.e., is the cascade time at $\kp \rp \sim 1$ the same as
it is in the MHD regime?).  Small uncertainties in C translate
directly (and exponentially) into large uncertainties in the energy
flux past $\kp \rp \sim 1$ and thus into large uncertainties in the
predicted electron heating rate.

It should be noted that equation (\ref{ekpa}) is only a crude
approximation for $E(\kp,\kpa)$.\footnote{For example, as noted above,
the GS cascade actually lies inside the cone $\kpa \lsim \omnl/v_A$,
not on the line $\kpa = \omnl/v_A$.}  Given the uncertainty in C,
however, alternative expressions do not give significantly different
results for the energy flux past $\kp \rp \sim 1$.

\section{Electron Heating Rate}
As before, we define $\delta$ to be the fraction of the turbulent
energy which heats the electrons; for the model in this paper, it is
given by $\delta \approx P_{TTD} + \epsilon/\epsilon_0$.  $P_{TTD}$ is
the (generally small) contribution to the electron heating from Alfv\'en
wave energy dissipated at $\kp \rp \lsim 1$.  $\epsilon/\epsilon_0$ is
the fraction of the turbulent energy which cascades past the damping
barrier at $\kp \rp \sim 1$ (essentially all of which heats the
electrons).  This is given by equation (\ref{flux}) of the previous
section.

Figure 2 shows our best estimates of $\delta$ as a function of $\beta$
for C = 24, 6, and 1.5 (from left to right).  For small $\beta$, the
electron heating is dominated by the energy that cascades past $\kp
\rp \sim 1$.  This is independent of the proton to electron
temperature ratio (since the mode dissipation rate at $\kp \rp \lsim
1$ is; see \S2), but is a strong function of $\beta$.  For larger
$\beta$ the dominant contribution to the electron heating is from the
dissipation of Alfv\'en waves at $\kp \rp \lsim 1$ (since almost no
energy cascades past $\kp \rp \sim 1$).  Since $P_{TTD} \approx (m_e
T_e / m_p T_p)^{1/2}$, this is a strong function of $T_p/T_e$, but is
independent of $\beta$.  In Figure 2, we have taken $T_p/T_e = 100$,
appropriate to accretion near the Schwarzschild radius of a black hole
(where most of the observed radiation originates).  Variations in
$T_p/T_e$ vertically shift the value of $\delta$ when it plateaus
(i.e., the high $\beta$ values), but do {\em not} significantly modify
the $\beta$ at which the plateau occurs.

From Figure 2 we infer that turbulence in ADAFs predominantly heats
the protons ($\delta \lsim 10^{-2}$, say) only for $\beta$ larger than
some critical value, which lies between $\sim 5$ and $\sim 100$ for
the values of C taken in Figure 2.  These values encompass what we
feel to be a reasonable estimate of the uncertainty in how the
turbulent energy cascades past $\kp \rp \sim 1$.  The corresponding
uncertainty in $\delta$ is extremely large, but this is an accurate
reflection of the (exponential) sensitivity of our results to unknown
details of turbulence on scales of the proton Larmor radius.

The conclusions to be drawn from Figure 2 depend, of course, on the
degree of one's confidence that all possible sources of electron
heating have been accounted for.  The most accurate interpretation of
Figure 2 is that it represents the fraction of the small scale
(Alfv\'enic) turbulent energy which heats the electrons.  In \S1, we
argued that this interpretation can be broadened to be (within a
factor of few) the fraction of both compressible and incompressible
turbulent energy which heats the electrons (since compressible
turbulence is unlikely to energetically dominate Alfv\'enic turbulence
and, in any case, does not significantly heat the electrons).

In Appendix B, however, we argue that, at large $\beta$, the physical
picture and calculations leading to Figure 2 neglect a source of
electron heating (possibly the most important one), namely magnetic
reconnection.  The reasoning is as follows.  Proton damping of the
turbulent energy at large $\beta$ is essentially a viscous dissipation
mechanism (since the protons carry the momentum of the plasma).  For a
nontrivial topology, the magnetic free energy of the turbulence cannot
be fully damped by viscosity.  This is because changes in the magnetic
topology, magnetic flux, magnetic helicity, etc., can only be due to
resistive effects (regardless of how small the resistivity is).  This
suggests that coincident with the dissipation of turbulent energy by
Alfv\'en wave damping is the formation of current sheets in which the
topology of the magnetic field changes.  Crude estimates suggest that
the energy dissipation in current sheets may be $\sim$ that of Alfv\'en
wave damping.  Unfortunately, we cannot assess whether $\sim$
corresponds to $1$ or $10^{-2}$, etc., and thus we cannot assess
whether reconnection is energetically important (that it is required
for topological changes is clear).  One way to address this question
is through numerical simulations of MHD turbulence with varying
magnetic Prandtl numbers (ratio of viscosity to resistivity).  The
relative contributions of Joule heating and viscous heating in such
simulations could potentially provide important information on the
energetic importance of magnetic reconnection in large magnetic
Prandtl number turbulence.

\section{Conclusions}
Spectral models of ADAFs generally assume that (see, e.g., Narayan et
al. 1998) (1) magnetic fields are amplified until they are in strict
equipartition with gas pressure ($\beta = 1$)\footnote{Narayan et al.
define the magnetic pressure as $B^2/24 \pi$ while the $\beta$ used in
this paper defines magnetic pressure as $B^2/8 \pi$.  Thus
equipartition for Narayan et al. actually corresponds to $\beta =
1/3$.}  and that (2) the energy generated by viscous stresses
predominantly heats the protons ($\delta \lsim 10^{-2}$ is a typical
value).

We suggest that these assumptions may be incompatible.  Based on an
analysis of incompressible turbulence (Alfv\'enic and whistler) and
collisionless wave dissipation, we find predominantly proton heating
only for $\beta$ greater than some critical value, which lies between
$\sim 5$ and $\sim 100$ depending on unknown details of how Alfv\'en
waves are converted into whistlers on scales of the proton Larmor
radius.  This does not, of course, imply that ADAF models are
untenable.  Rather, their sensitivity to changes in input microphysics
(e.g., $\beta$ and $\delta$) should be carefully assessed.  This will
be pursued in a future paper.

We cannot overemphasize the uncertainty in the numerical values given
in this paper (which are, e.g., based on the assumption of a uniform
thermal plasma).  Nonetheless, we believe that the basic physical
picture of particle heating (wave damping + reconnection), and the
general conclusions drawn from it ($\delta$ increases as $\beta$
decreases), are essentially correct.  One point which clearly requires
additional investigation is the energetic importance of magnetic
reconnection (\S5 and Appendix B).

\noindent{\it Acknowledgments.}  EQ thanks George Field, Charles
Gammie, Jason Maron, and Ramesh Narayan for useful discussions.  EQ
was supported by an NSF Graduate Research Fellowship and, in part, by
NSF Grant AST 9423209. AG thanks Jason Maron, Peter Goldreich, and NSF 
PHY-95-13835.

\newpage

\begin{appendix}

\section{Whistler Turbulence}
The purpose of this appendix is three-fold.  First, we argue that
Alfv\'enic turbulence is converted into whistler turbulence on scales of
the proton Larmor radius.  While the details of this process are
uncertain, all of the turbulent energy not damped from Alfv\'en waves on
scales of $\sim \rp$ is transformed into whistler energy on smaller
scales (there are no other channels through which the energy can
travel).  Second, we discuss the turbulent cascade of whistlers.
Finally, we discuss the dissipation of whistler turbulence.  This,
together with \S2 on Alfv\'enic turbulence, provides a picture (if
approximate and uncertain in detail) of turbulence and particle
heating from the outer to the inner scale.

\subsection{Alfv\'en $\rightarrow$ Whistler:  $\kp \rp \sim 1$}
For $\beta \sim 1$, the full kinetic theory dispersion relation
evolves continuously from Alfv\'en waves to whistlers as we pass $\kp
\rp \sim 1$.  This is because whistlers are the natural generalization
of Alfv\'en waves once protons drop out of the small-scale dynamics due
to Larmor circle averaging.  Consequently, Alfv\'en waves at $\kp \rp
\lsim 1$ can excite whistlers at $\kp \rp \gsim 1$ by three-wave
interactions (i.e., the resonance conditions can be
satisfied).\footnote{Note that because the GS cascade is strong, the
frequency ``resonance'' condition is actually quite broad.}  For
$\beta \gsim 100$, the situation is more complicated because there is
a region (the ``damping barrier'') of wavevector space ($ 0.85 \lsim
\kp \rp \lsim 1.15$ for $\beta \sim 100$ and larger for larger
$\beta$) in which Alfv\'en waves do not propagate (see G98).  Because
this region is rather narrow for the $\beta$ of interest to us, Alfv\'en
waves can excite whistlers across the damping barrier (narrow meaning
that the jump in $\kp$ is less than a factor of $\sim 2$).  This
establishes that the turbulent energy not damped on scales of $\kp \rp
\sim 1$ can be converted to whistler energy on smaller scales
(although, as discussed in \S4, the precise details -- e.g., the
timescale -- are uncertain).

To establish that no other channels of energy travel are possible is
straightforward.  Whistlers are the only modes with $\kp \rp \gsim 1$
and $\om \ll \Omega_p$ (``sound waves'' are too strongly damped to be
excited in collisionless plasmas; see Q98).  They are therefore the
only possible sink of the Alfv\'enic energy not damped on scales of $\kp
\rp \sim 1$.

\subsection{Whistler Cascade}
Like collisionless Alfv\'en waves, collisionless whistlers can be
described hydrodynamically. Hydrodynamic equations valid at $k_{\perp
}\rp \gtrsim 1$ can be obtained as follows. The protons are
dynamically frozen; their sole function is to create a positive charge
background on which the electrons and magnetic fields evolve.
Electrons move freely along magnetic field lines and so
$E_{\parallel}=0$.  In the perpendicular direction, electrons move
with the E$\times $B-drift velocity, ${\bf v}_{\perp }=c{\bf E}\times
{\bf B}/B^2$. These two equations give
\begin{equation}
{\bf E}+{1\over c}{\bf v}\times {\bf B}=0. \label{ep}
\end{equation}
Neglecting displacement currents, 
\begin{equation}
\nabla \times {\bf B}=-{4\pi \over c}ne{\bf v}.\label{fd}
\end{equation}
From equations (\ref{ep}) and (\ref{fd}), we obtain the Electron
Magnetohydrodynamics (EMHD) equation (Kingsep, Chukbar, \& Yan'kov
1990)
\begin{equation}
\partial _t {\bf B}={c\over 4\pi ne}\nabla \times ({\bf B}\times
\nabla \times {\bf B}). \label{EMHD}
\end{equation}
The linear waves in equation (\ref{EMHD}) are whistlers; their
dispersion relation can be presented as
\begin{equation}
\omega ={v_A^2\over v_p}|(k_{\perp }\rp )k_{\parallel }|, \label{whist}
\end{equation}
where $v_p$ is the proton thermal speed.

To see at what scales the protons freeze out, we compared the whistler
dispersion relation obtained numerically from a full plasma
permittivity tensor (including both electrons and protons) to the
analytical dispersion relation (\ref{whist}). At $\beta =100$, the
dispersion relations agree to 30 \% at $k_{\perp }\rp=1.5$, to 10 \%
at $k_{\perp }\rp=2$, and to 3 \% at $k_{\perp }\rp=3$.\footnote{Other
relevant plasma parameters in this example are $v_e/c=0.5$,
$v_p/c=0.33$, $k_{\parallel }\ll k_{\perp }$. For details see Q98 or
G98.}

The turbulent cascade in (\ref{EMHD}) was described by Kingsep,
Chukbar, \& Yan'kov (1990).  A very clear discussion is given by
Goldreich \& Reisenegger (1992). These authors assumed that whistler
turbulence is roughly isotropic in wavenumber space, and that the
turbulence is weak. Their argument for the turbulence being weak is
analogous to the standard argument for the ``weakness'' of Alfv\'enic
turbulence. Assuming strong isotropic turbulence, one calculates a
nonlinear frequency at a given scale.  This frequency turns out to be
smaller than the linear wave frequency; the turbulence must therefore
be weak.

As explained by GS for the MHD case, however, the above
``explanation'' does not necessarily work. Instead of being weak and
isotropic, the turbulence is strong in a narrow cone in wavenumber
space ($\kpa \ll \kp$), for which the linear frequency is smaller than
the nonlinear frequency. Numerical simulations seem to confirm the GS
picture in the Alfv\'en wave case (Maron 1998). But the same picture
must hold true (to some extent) in the whistler case.  In a narrow
cone in wavenumber space, the turbulent cascade can proceed at a
(fast) nonlinear rate.  Outside of the cone, the cascade slows down
(the turbulence is weak). This is because nonlinear interactions
outside the cone must satisfy the frequency resonance condition
(interactions are restricted to the resonant manifold).

For our problem, energy is injected into whistler turbulence in the
narrow cone for which the turbulence is strong (because the Alfv\'enic
turbulence which excites whistler turbulence is itself strong).  This
suggests that a large part of the energy will continue to cascade in
the strong cone, maintaining $k_{\parallel} \ll k_{\perp}$. Numerical
simulations in two dimensions (Biskamp, Schwarz \& Drake 1996) confirm
that strong Kolmogorov turbulence results in both the Alfv\'en and the
whistler cases.  

We have argued that the energy in whistler turbulence injected at
$k_{\parallel} \ll k_{\perp}$ preserves this anisotropy because the
turbulent energy cascade is faster in the strong cone where the
nonlinear frequency shifts exceed the linear frequency.  We cannot,
however, rule out that the strong cone may be leaky. It is a peculiar
feature of Alfv\'enic turbulence, not shared by its whistler
counterpart, that the 3-wave (and 4-wave) resonance conditions drive
the turbulent energy into the strong cone, regardless of their
starting point in wavevector space.  Our analysis of whistler
turbulence is therefore only suggestive.  Thankfully, the precise path
of whistler turbulence in wavevector space is not needed to assess its
particle heating properties.  All that is needed is that it maintains
$\kpa \ll \kp$ (this is quantified below).

\subsection{Whistler Dissipation}
In this subsection, we assess the dissipation of whistler turbulence.
Assuming that whistler turbulence
maintains $\kpa \ll \kp$, we show that the protons cannot be heated by
whistlers.

For a (subthermal) wave with $\kp \rp \gsim 1$ damped solely by
transit time damping, the relative heating of electrons and protons is
given by \beq P_{TTD} \simeq \left(m_e T_e \over m_p T_p \right)^{1/2}
\left(\kp \rp \right)^3
\label{TTD}.  \eeq
The $(\kp \rp)^3$ factor describes quantitatively the effects of
Larmor circle averaging (compare with the $\kp \rp \lsim 1$ expression
in \S2).  For $T_p \sim 100 T_e$, electrons are heated more than
protons for $\kp \rp \gtrsim 7$.\footnote{Note that for all $\kp \rp$
Landau damping preferentially heats the electrons for $T_p \gg T_e$.}
The decrease in the whistler dissipation rate for $\kp \rp \gsim 1$ in
Figure 1 is due to the freezing out of the protons.  Even though the
protons are in principle heated more than the electrons for $\kp \rp
\lsim 7$, the decrease in the dissipation rate ensures that very
little energy is dissipated between $1 \lsim \kp \rp \lsim 7$.  For
length scales smaller than $\kp \rp \sim 7$, Larmor circle averaging
entails that protons cannot be efficiently heated by the $n = 0$
Landau resonance (see eq. [\ref{res}]).  This conclusion is
essentially independent of the details of the whistler cascade.

For $\kp \rp \gsim 1$, protons can also be heated by the cyclotron
resonance (for wave frequencies $\sim$ the proton cyclotron frequency;
the $|n| = 1$ resonance in eq. [\ref{res}]).  For large $\kp \rp$, the
dominant electron heating mechanism is Landau damping.  The relative
importance of proton heating by the cyclotron resonance and electron
heating by Landau damping is given roughly by $(m_p/m_e) (\kp
\rp)^{-3}$, where we have taken $v_A \sim v_p \sim v_e \sim v \sim c$
and $\omega \sim \Omega_p$.\footnote{In deriving this expression we
have used an analytical approximation for the relative strengths of
the perpendicular and parallel electric field components of the
whistler wave (from G98).}
  
For $\kp \rp \gsim 10$, Larmor circle averaging is sufficiently strong
that proton heating by the cyclotron resonance is less important than
electron heating by Landau damping.\footnote{Our numerical
calculations with the full plasma permittivity tensor confirm this
result.}  Only if whistler turbulence reaches $\omega \sim \Omega_p$
before $\kp \rp \sim 10$ can the protons be heated by whistler
turbulence.  This would require a complete reversal in the direction
of the turbulent cascade since at $\kp \rp \sim 1$ the GS cascade
yields $\omega \sim 10^{-3} \Omega_p$.  Our estimates of whistler
turbulence in the previous subsection suggest that it reaches $\omega
\sim \Omega_p$ only for $\kp \rp \gg 10$.  Consequently, Larmor circle
averaging of the protons is efficient and the whistler energy is
damped by the electrons (by Landau damping) on scales of $\kp \rp \sim
30$ (see Figure 1).

\section{Reconnection}
In this appendix, we discuss why reconnection (Joule heating) may be
the primary heating mechanism for the electrons, especially at low
magnetic fields (when whistlers are not significantly excited). We
argue that, to order of magnitude, comparable amounts of energy are
damped by viscous heating of the protons and by Joule heating of the
electrons.  Bisnovatyi-Kogan \& Lovelace (1997) first suggested that
Joule heating of the electrons could be important in ADAFs (we do not,
however, agree with the particular scenario they propose; see Q98 or
Blackman (1998)). Blandford (1998) also suggested that reconnection
could be important.  The following analysis attempts to assess the
potential significance of reconnection in a semi-rigorous manner.

For $\beta \gg 1$, Alfv\'en waves are damped at $k_{\perp}\rp <1$ and
the plasma dynamics can be described by the MHD equations. There are
two heating mechanisms in the framework of MHD - viscous heating and
Joule heating. Roughly speaking, viscosity ($\nu$) heats the particles
that carry the momentum, i.e., the protons, and resistivity ($\eta$)
heats the particles that carry the current, i.e., the electrons. We
know that $\nu \gg \eta$ in a $\beta \gg 1$ plasma - protons are
heated when Alfv\'en waves are damped by viscosity.

Is it true, however, that Joule (electron) heating dissipates much
less energy than viscous (proton) heating in a $\nu \gg \eta$ plasma
(i.e., for large magnetic Prandtl numbers)?  In other words, can
Alfv\'en wave damping actually dissipate all of the turbulent energy?

The magnetic field of the accreting plasma is ostensibly
self-generated.  The magnetic field energy is constantly being
created, destroyed, and created again. To demonstrate that
reconnection is unavoidable, consider the relaxation of a magnetic
field perturbation created ``by hand'' at time zero. This initial
magnetic field is assumed to be a generic solenoidal field with a
characteristic strength $B_0$ and a characteristic scale $L$; we also
assume that the normal component of the field is specified at the
boundary of a box with side $\sim L$. Now let the field evolve
according to MHD equations with $\nu \gg \eta$.

Magnetic pressure and tension cause plasma motions on scales $\sim L$.
A Kolmogorov-GS cascade develops and magnetic energy is converted into
kinetic energy; viscous heating by the protons dissipates the kinetic
energy. In a time $\sim L/v_A$, the kinetic energy goes to zero and
the magnetic field reaches an equilibrium state defined by
\begin{equation}
\nabla \times {\bf B}\times {\bf B}=4\pi \nabla p, 
\end{equation}
where $p$ is the plasma pressure. Generically $B^2\sim B_0^2\sim
(B_0-B)^2$ so that the energy heating the protons is $\sim B_0^2L^3$.

As explained by Arnold (1986), the equilibrium magnetic field lines
described by equation (B1) have a very special topology. They lie on a
set of nested tori with $p={\rm const}$.  For $\eta = 0$, however, the
magnetic field is frozen into the plasma (magnetic flux, helicity, and
the topology of the field lines are all conserved). Since the initial
magnetic field topology was a generic one (by assumption), it is
impossible to reach the equilibrium configuration with $\eta = 0$
(since that requires changing the topology of the field). On the other
hand, the equilibrium must be reached, because out of equilibrium the
magnetic field causes plasma motions and therefore viscous heating.

The resolution of the paradox is the formation of current sheets.
Topological barriers on the way to equilibrium are squeezed to zero
volume. No matter how small, resistivity becomes important in the
current sheets thus allowing reconnection ($\equiv$ changes in the
magnetic field topology). Generically the magnetic free energy
released at this stage is $\sim B_0^2L^3$.

The only way to avoid reconnection is to assume that the magnetic
field has a trivial topology. For example, one can assume that
magnetic field lines constantly lie on surfaces that are topologically
equivalent to a set of nested tori.  The dynamo nature of the magnetic
field in accretion flows, however, seems incompatible with a trivial
topology ($cf$ the complex structures observed in numerical
simulations of the Balbus-Hawley instability).

While we do not know how to rigorously estimate the fraction of the
turbulent energy damped in reconnection events, the potential
importance of reconnection for electron heating, and its
unavoidability even in large magnetic Prandtl number plasmas, seems
clear.

\end{appendix}

{
\footnotesize
\StartRef
\noindent {\large \bf References} \\
\Ref Abramowicz, M., Chen, X., Kato, S., Lasota, J. P, \& Regev, O., 1995,
ApJ, 438, L37 \\
\Ref Achterberg, A. 1981, A\&A, 97, 259 \\
\Ref Arnold, V. I. 1986, Sel. Math. Sov., 5, 327 \\
\Ref Balbus, S. A., \& Hawley, J. F. 1991, ApJ, 376, 214 \\
\Ref Biskamp, D.,  Schwarz, E., \& Drake, J. F.  1996, Phys. Rev. Lett., 76, 
1264 
\\
\Ref Bisnovatyi-Kogan, G. S., \& Lovelace R. V. E 1997, ApJ Lett., 486, 43 \\
\Ref Blackman, E. 1998, Phys. Rev. Letters, in press (astro-ph/9710137)\\
\Ref Blandford, R. D. 1998, preprint \\
\Ref Goldreich, P., \& Reisenegger, A. 1992, ApJ, 395, 250 \\
\Ref Goldreich, P. \& Sridhar, S. 1995, ApJ, 438, 763 (GS)\\
\Ref Gruzinov, A. 1998, ApJ in press (astro-ph/9710132) (G98)\\
\Ref Ichimaru, S. 1977, ApJ, 214, 840 \\
\Ref Kingsep, A. S.,  Chukbar, K. V., \& Yan'kov V. V. 1990, Rev. Plasma Phys., 16, 24 \\
\Ref Maron, J. 1998, private communication \\
\Ref Narayan, R., Mahadevan, R., \& Quataert, E., 1998, in preparation \\
\Ref Narayan, R., \& Yi, I., 1994, ApJ, 428, L13 \\ 
\Ref Narayan, R., \& Yi, I., 1995, ApJ, 452, 710 \\ 
\Ref Quataert, E. 1998, ApJ in press (astro-ph/9710127) (Q98)\\
\Ref Rees, M. J., Begelman, M. C., Blandford, R. D., \& Phinney,
E. S., 1982, Nature, 295, 17 \\
\Ref Stix, T.H. 1992, Waves in Plasmas (New York: AIP) \\ 
}

\newpage  

\begin{figure}
\plotone{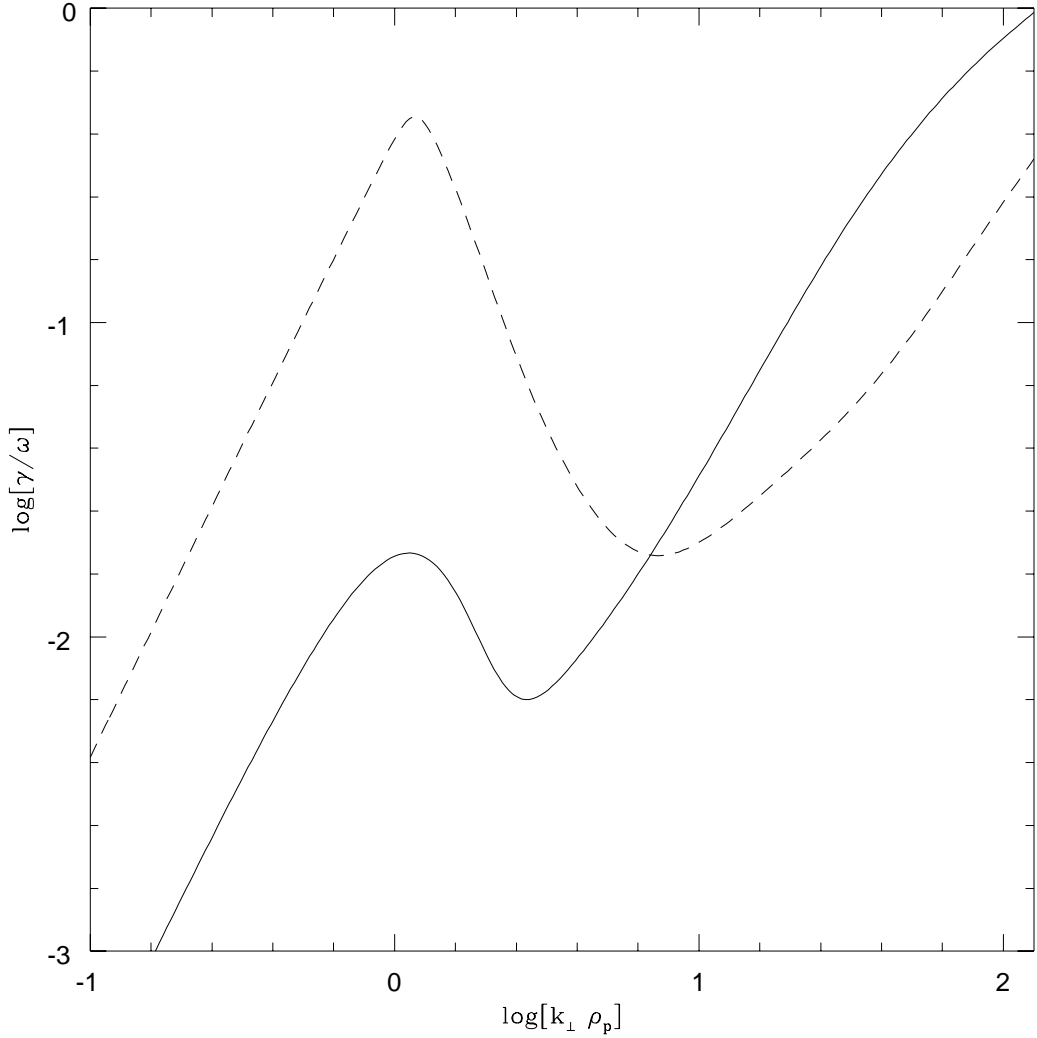}
\caption{Dimensionless dissipation rate of the Alfv\'en ($\kp \rp <
1$) and whistler ($\kp \rp > 1$) modes, taking $\kpa \ll \kp$.  The
solid (dashed) curve is for $\beta = 1 (10)$, while both curves take
$T_p = 100 T_e$.  The peak in the dissipation at $\kp \rp \sim 1$
corresponds to proton heating while the strong damping at $\kp \rp >
30$ corresponds to electron heating.}
\end{figure}

\newpage

\begin{figure}
\plotone{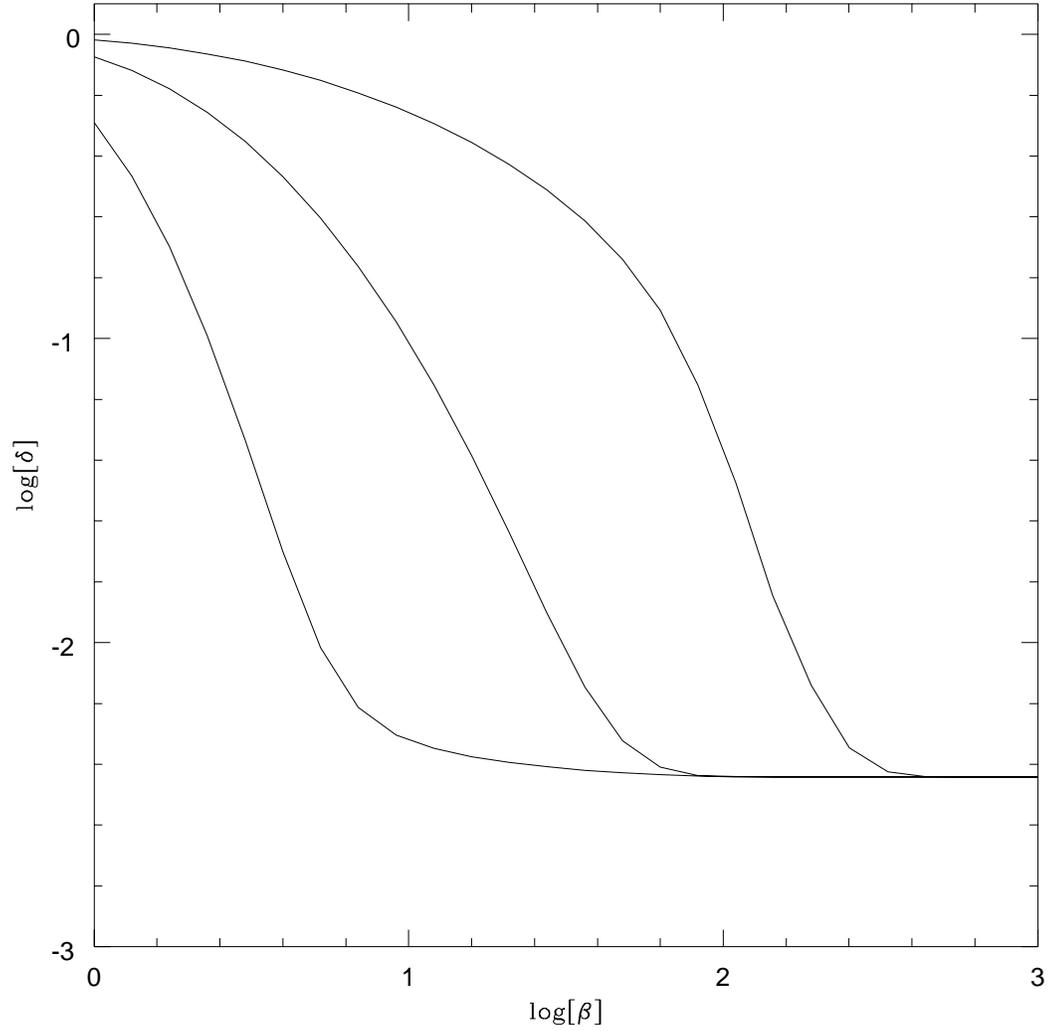}
\caption{Estimates of $\delta$, the fraction of the turbulent energy
which heats the electrons, versus $\beta$ (for $T_p = 100 T_e$).  The
three curves correspond to (from left to right) C = 24, 6, and 1.5
(see eq. [\ref{flux}]).}
\end{figure}

\end{document}